# Induction of microRNAs mir-155, mir-222, mir-424 and mir-503, promotes monocytic differentiation through combinatorial regulation.

Alistair R. R. Forrest*[1,2], Mutsumi Kanamori-Katayama[1], Yasuhiro Tomaru[1], Timo Lassmann[1], Noriko Ninomiya[1], Yukari Takahashi[1], Michiel J. L. de Hoon[1], Atsutaka Kubosaki[1], Ai Kaiho[1], Masanori Suzuki[1], Jun Yasuda[1], Jun Kawai[1], Yoshihide Hayashizaki[1], David A. Hume[3], Harukazu Suzuki[1]

[1]Omics Science Center, RIKEN Yokohama Institute, 1-7-22, Suehiro-cho, Tsurumi-ku, Yokohama,230-0045, Japan

[2]The Eskitis Institute for Cell and Molecular Therapies, Griffith University, Brisbane, QLD 4111, Australia.

[3]The Roslin Institute and Royal (Dick) School of Veterinary Studies, The University of Edinburgh, Roslin, EH259PS, UK.

**Corresponding Author:** Dr Alistair Forrest, Omics Science Center, RIKEN Yokohama Institute, 1-7-22 Suehiro-cho, Tsurumi-ku, Yokohama, Kanagawa 230-0045 Japan. Fax: +81-45-503-9216; Tel:+81-45-503-9222; Email: alistair.forrest@gmail.com

# Abstract

Acute myeloid leukemia (AML) involves a block in terminal differentiation of the myeloid lineage and uncontrolled proliferation of a progenitor state. Using phorbol myristate acetate it is possible to overcome this block in THP-1 cells (a M5-AML containing the MLL-MLLT3 fusion), resulting in differentiation to an adherent monocytic phenotype. As part of FANTOM4 we used microarrays to identify 23 microRNAs that are regulated by PMA. We identify four PMA induced microRNAs (mir-155, mir-222, mir-424 and mir-503) that when over-expressed cause cell cycle arrest and partial differentiation and when used in combination induce additional changes not seen by any individual microRNA. We further characterise these pro-differentiative microRNAs and show that mir-155 and mir-222 induce G2 arrest and apoptosis respectively. We find mir-424 and mir-503 are derived from a polycistronic precursor mir-424-503 that is under repression by the MLL-MLLT3 leukemogenic fusion. Both of these microRNAs directly target cell cycle regulators and induce G1 cell cycle arrest when over-expressed in THP-1. We also find that the pro-differentiative mir-424 and mir-503 down-regulate the anti-differentiative mir-9 by targeting a site in its primary transcript. Our study highlights the combinatorial effects of multiple microRNAs within cellular systems.

# Introduction

The FANTOM4 (Functional ANnoTation Of Mammals) project used a combination of deep sequencing, microarrays, bioinformatic predictions and siRNA perturbations to map a network of mammalian transcription factors and their targets (1). For this project we studied monocytic differentiation of the M5 acute myeloid leukemia (AML) cell line THP-1 (2). This cell line contains the MLL-MLLT3 leukemogenic gene fusion (3) and displays a monoblastic phenotype, however upon phorbol myristate acetate (PMA) treatment it differentiates into an adherent monocytic phenotype (4), closely approximating monoblast to monocyte differentiation.

In this parallel study we monitored the expression dynamics of microRNAs after PMA treatment and sought to identify those that might promote differentiation. MicroRNAs are short 21-22 nucleotide RNA molecules which cause translational repression and degradation of multiple messenger RNAs by binding target sequences in their 3' UTRs (5). Multiple microRNAs are implicated in differentiation of cell lineages including skeletal muscle, adipocytes and neurons (6-8) and in mammals, the microRNA sub-system is essential for embryonic development. Dicer is required for differentiation of embryonic stem cells (9), Dicer (-/-) knockouts are embryonic lethal (10) and Ago2 (-/-) knockouts suffer severe haematopoietic defects (11). In addition several microRNAs are deleted or amplified in cancer and have been implicated as tumour suppressors (12, 13) or oncogenic (14, 15).

In this study we find multiple microRNAs are induced upon PMA treatment and that four of these can promote both partial monocytic differentiation and cell cycle arrest in over-expression experiments. Our data suggests a model whereby multiple microRNAs are induced in conjunction with various transcription factors to co-operatively promote differentiation and inhibit cellular proliferation at multiple points in the cell cycle.

## Materials and methods

The details of this section are available online as Supplementary Information.

## Results

**Multiple microRNAs are induced during PMA induced THP-1 differentiation.**

The expression patterns of microRNAs during PMA induced THP-1 differentiation were determined using Agilent microRNA arrays on biological triplicate time-courses. The expression profile for each microRNA after PMA treatment is available through the FANTOM4 EdgeExpress Database (16). Between undifferentiated (0h) and differentiated (96h) states we identified 21 microRNAs that were up-regulated and 2 that were down-regulated with average expression changes of at least 3 fold (Table 1). Of the up-regulated microRNAs, several have been previously implicated in haematopoiesis including mir-424 which promotes monocytopoiesis (17), mir-221, mir-222 and mir-24 which inhibit eryrthopoiesis (18, 19) and mir-155 which promotes B-cell proliferation and depletes myeloid and erythroid hematopoietic stem cell populations (20, 21). Of the two down-regulated microRNAs, mir-210 promotes osteoblast differentiation (22) and mir-9 blocks B-cell differentiation by targeting PRDM1 (23). We note that although detected in THP1 the mir-17-92 polycistron previously reported to prevent monocytic differentiation by repression of AML1 (24) is not significantly down-regulated suggesting our cells are at a different stage of differentiation.

Deep-sequencing of small RNAs from one of the three THP-1 biological replicates time-courses (25) confirmed the differential expression of microRNAs detected by the arrays. Fold change estimates by the two technologies varied (eg. mir-221 was induced 5.7 fold on the arrays but 10.9 fold by sequencing) but repression or induction was consistent. We also estimated microRNA abundances from the sequencing and found mir-221 the most abundant inducible microRNA in THP-1 accounting for 11% of all tags at 96h (106251 tags per million).

**MicroRNAs 155, 222, 424 and 503 partially promote monocytic differentiation**

A set of 11 PMA regulated microRNAs (10 induced and 1 repressed) with a wide range of abundances (eg. mir-146b – 23 tpm, mir-221 – 106251 tpm) were selected for further studies. Synthetic microRNA precursors were transfected into undifferentiated THP-1 cells and the effect on gene expression measured after 48hrs using Illumina Sentrix6v2 whole genome microarrays. Significant expression changes were determined by B-statistic comparison (26) against the effect of a scrambled negative control duplex RNA (B-statistic ≥ 2.5 and fold change ≥ 2) (Supplementary table 1). Lists of up and down-regulated mRNAs were then compared to PMA regulated mRNAs identified in the FANTOM4 data (1) to identify microRNAs that promote differentiation. Expression changes induced by the microRNA over-expressions could be then split into PMA-like (pro-differentiative) changes, unrelated changes and anti-PMA-like (anti-differentiative) changes.

MicroRNAs were considered pro-differentiative if 1) there were more PMA-like changes than anti-PMA-like changes (at least 1.5x) and 2) for the given microRNA the fraction of all changes that were PMA-like was greater than the average fraction plus one standard deviation (Supplementary Table 2). None of the microRNAs completely reiterated the expression changes observed with PMA, and no morphological differences were observed, however using the above criteria mir-155, mir-222, mir-424 and mir-503 induced expression changes suggesting partial differentiation including induction of known markers of differentiation ITGAL (Cd11a), ITGB7 and ENG (Cd105) by mir-155 and ITGAL, ITGB2 and CSF1R by mir-503 (Supplementary Fig. 1). Mir-155 and mir-503 had the greatest ratio of PMA-like to anti-PMA-like changes (2.9x) and mir-503 also had the greatest number of PMA-like changes in total (75 down, 35 up). As a comparison the additional negative control microRNA (mir-142 – expressed in THP-1 but not regulated by PMA) failed to induce pro-differentiative expression changes and had a differentiation ratio of 0.3x.

The combined effect of the four microRNAs was then tested by co-over-expressing mir-155, mir-222, mir-424 and mir-503 in equimolar amounts in THP-1 for 48hrs. The mixture induced 69 pro-differentiative changes, 39 of which were not observed in any of the four individual pre-miRNA transfections (including modest induction of the key monocytic marker CD14 - supplementary figure 2). One quarter of all changes induced by the mixture were PMA like and the ratio of PMA-like to anti-PMA-like changes was 6.3x, far larger than observed for any of the individual miRNA transfections (Supplementary Table 2). Together this suggests that the mixture more closely mimics the PMA induced expression changes by reducing the non-specific and anti-PMA-like changes and increasing the PMA-like changes.

Finally we also identify mir-9 as an anti-differentiative miRNA using this analysis. 36% of the genes down-regulated by mir-9 over-expression are normally up-regulated with PMA and 26% of all expression changes for mir-9 were anti-differentiative suggesting it helps maintain the undifferentiated monoblast state similar to its role in blocking B-cell differentiation (23).

**mir-424-503 is a polycistronic microRNA cluster that promotes differentiation and G1 arrest of the cell cycle.**

Gene ontology analysis (27) of genes affected by the pro-differentiative microRNAs revealed that both mir-424 and mir-503 down-regulate genes annotated as cell-cycle regulators (Table 2 and supplementary table 3). This is consistent with a report that mir-424 over-expression can induce G1 arrest of the cell cycle (28). We confirmed this in THP-1 cells using flow cytometry and also show that mir-503 induces G1 arrest to similar levels as mir-424 (Fig. 1). In addition we find mir-222 induces a G2 accumulation, while mir-155 was depleted for G2 with accumulation of a sub-G1 population, suggesting apoptosis.

Examination of the mir-424 and mir-503 loci finds they are separated by 383 bases on the genome and likely to be derived from the same primary transcript. We confirm this using RT-PCR (Fig 2A & B, Supplementary Fig.4). A further five microRNAs (mir-542-5p, mir-542-3p, mir-450a, mir-450b-5p, mir-450b-3p) that are within 7KB of mir-424-503 may also be generated from the same primary transcript, and two of these (mir-542-5p and mir-542-3p) are also induced. However the deep sequencing results show these are present at much lower levels than mir-424 and mir-503 (see Table 1).

Finally we observe that the mir-424-503 polycistronic microRNAs have related seed sequences, thus they share target genes explaining why both can induce partial differentiation and cell cycle arrest (Fig. 2C). The array analysis shows overlapping but distinct sets of targets for both mir-424 and mir-503, and despite similarity to mir-15/16 members neither induced apoptosis (29). This arrangement of polycistronic microRNAs with shared seed sequences has been reported for mir-15a-mir-16-1 (29, 30), and mir-17-92 (14), and may be a common architecture.

**mir-424 and mir-503 directly target multiple cell cycle regulators.**
Mir-424 has been previously shown to target the cell cycle regulators CCNE1, CCND1, CCND3 and CDK6 (28, 31) and all of these are down-regulated at the mRNA level in our over-expression experiments (0.77, 0.84, 0.69 and 0.85 fold respectively). To identify the likely direct targets of mir-503 and the other microRNAs identified in this study we compared publicly available microRNA target predictions (both EIMMO and TargetScan (32, 33)) with the lists of genes down-regulated in the microRNA over-expression experiments (for full lists of putative targets see supplementary table 4 and also supplementary Fig. 3).

A combined set of predicted mir-424 and mir-503 targets with roles in cell cycle were then tested using luciferase reporter assays with mir-424, mir-503 and mir-9 co-transfection (Fig. 3). The known mir-424 targets CCNE1 and CCND1 were the second and third most affected targets of mir-424 and were also affected by mir-503, while PRDM1 and ETS1 which lack predicted mir-424/503 sites were the least affected (Fig. 3A). Conversely PRDM1 was the most affected construct in the mir-9 transfection, while WEE1 and CDC14A lacking mir-9 target sites were the least affected.

Using BCL6 and POU2F2 constructs as thresholds for non-specific down-regulation (these both lack mir-424/503 sites) we find the following, mir-424 targets ANLN, CCNE1, CCND1, WEE1, ATF6, KIF23, CHEK1, CDC25A, CDC14A and CCNF while mir-503 targets CDC14A, ANLN, CCND1, ATF6, EIF2C1, CDC25A, CHEK1, CCNE1, CCNE2, WEE1, CCNF and CDKN1A. We note that seven of these targets appear in both lists, however they are affected to different levels eg. CDC14A was the most affected target of mir-503 but tenth strongest by mir-424, reflective of the different affinities of the mir-424 and mir-503 sequences.

**The pro-differentiative mir-424-503 polycistron also targets the primary transcript of the anti-differentiative mir-9-3**

Mentioned previously, mir-9 is down-regulated during differentiation and when over-expressed induces anti-differentiative gene expression changes. The mature microRNA can be generated from three possible locations, mir-9-1, 9-2 and 9-3, however mir-9-3 is the most likely copy regulated in THP-1 differentiation as both Illumina microarray signal (Fig. 4A) and Cap Analysis of Gene Expression signal (data not shown) for the primary transcript is down-regulated with PMA. Surprisingly, over-expression of either mir-424 or mir-503 also down-regulated mir-9-3 (Fig. 4A) and upon examining the primary transcript sequence we identified a potential mir-424-503 target sequence (Supplementary figure 5). When tested in a luciferase assay the sequence was targeted by both mir-424 and mir-503 (underlined Fig. 3). This suggests that at least *in vitro* the pro-differentiative microRNAs mir-424 and mir-503 may directly down-regulate the anti-differentiative mir-9.

**Regulation of the mir-424-mir-503 polycistron**

A review of public expression data for the primary transcript (MGC16121) finds that it is serum inducible in fibroblasts (34) and up-regulated during normal monocyte to macrophage maturation (35) (Gene Expression Omnibus (GEO) (36) entries: GDS1568 / 8259, 16661, 40711 / MGC16121 and GDS2430 / 227488_at, 229784_at / MGC16121). In addition, from the FANTOM4 knockdown panel of 52 myeloid transcription factors (1) we find knock-down of the MLL-MLLT3 leukemogenic fusion of THP-1 induced mir-424-503 (Fig. 4B). Knockdown of MLL-MLLT3 has previously been shown to induce differentiation and cessation of cellular proliferation (37, 38). Our analysis suggests that repression of mir-424-503 by MLL-MLLT3 may help maintain proliferation and prevent cell cycle arrest and differentiation by mir-424 and mir-503, two putative tumour suppressor microRNAs of the mir-15/16 family (29, 39).

## Discussion

Twenty three microRNAs were identified that are dynamically expressed during monocytic differentiation of the THP-1 acute myeloid leukemia cell line. Four of these induce partial differentiation and cell cycle arrest when over-expressed. Three have been previously implicated in haematopoietic differentiation (17, 19, 21) however this is the first report implicating mir-503 and suggesting they co-operate to promote monocytic differentiation. Mir-424 was previously reported as regulated by SPI1(PU.1) and inducing monocytic differentiation by targeting NFIA (17), however in THP-1, NFIA is not expressed and SPI1 knockdown has no effect on mir-424 expression suggesting other targets and regulators are involved. We show that mir-424 and mir-503 are produced as a polycistronic message which is repressed by MLL-MLLT3 and that both induce G1 arrest by targeting an overlapping set of cell cycle regulators. We also show both target the primary transcript of mir-9 an anti-differentiative microRNA. This is the first report to our knowledge of a set of pro-differentiative microRNAs targeting the primary transcript of an anti-differentiative microRNA.

For mir-155 and mir-222 we observed apoptosis and G2 cell cycle arrest respectively. From the array data and target site predictions we identify several targets that could explain the respective phenotypes (see supplementary table 4). We hypothesize that apoptosis by mir-155 occurs by targeting anti-apoptotic factors (similar to BCL2 targeting by mir-15/16 (29)) and predict RPS6KA3, SGK3, RHEB and KRAS as likely targets (40-43) three of which have supporting protein level evidence as targets (RPS6KA3, RHEB and KRAS) (44, 45). Apoptosis by mir-155 has not been previously reported but may explain the selective depletion of myeloid and erythroid hematopoetic stem cell populations in preference for B-cell proliferation (21), and also explain the fraction of apoptotic THP-1 cells observed with PMA treatment (46).

We also predict the G2 accumulation of THP-1 induced by mir-222 is via targeting of CDKN1B, RB1 and SKP1A by mir-222. Knock-down of RB1 and SKP1A can increase G2 populations (47, 48) and CDKN1B is a validated target (49, 50). We note that in different cell lines mir-222 can induce either an S-phase or G1-accumulation (51, 52). This highlights that the effect of mir-222, like mir-155 and perhaps most microRNAs is context specific. The phenotype in any given cell type will depend on the expression levels of microRNA targets and all the other components with which they are networked.

The interconnected nature of these components is best demonstrated by the observations regarding the four microRNAs. All four were shown to induce cell cycle phenotypes and for mir-424 and mir-503 we confirmed down-regulation of <u>direct</u> targets by luciferase assay, however these microRNAs also induced significant numbers of genes. For example 57 of the 209 genes induced by the four microRNA mix are also induced by PMA. These must be <u>indirect</u> targets, which are induced through down-regulation of anti-differentiative components of the monoblast state, such as pri-mir-9-3 (Fig. 5).

Finally THP-1 differentiation involves the co-ordinated up-regulation of transcription factor activities involved in the differentiated monocyte phenotype and down-regulation of factors primarily involved in cell cycle (1). Artificial siRNA knockdown of these naturally down-regulated factors or over-expression of the four microRNAs individually can promote partial differentiation however none are sufficient to completely reiterate differentiation in THP1. This is most likely because state transition (from one 'transcriptional basin' (1) to another) requires co-ordination of multiple components including transcription factors, microRNAs and cytokines, and perturbation of a single component is insufficient to perturb the system from one stable basin state to another. To test this in part we have shown that co-over-expression of all four microRNAs promotes a more specific PMA like response than over-expression of any individual microRNA, and it would be interesting to combine this further with knock down of specific transcription factors such as MYB and MLL-MLLT3. The data we have presented here has been generated using the THP-1 cell line containing the MLL-MLLT3 fusion. It will be important now to test the role of these microRNAs in primary AML samples and cell lines with other known genetic legions and compare them with normal monoblasts to see whether over-expression of these microRNAs could be used as a potential differentiative therapy.

## Acknowledgements

This study was supported by the following; A research grant for RIKEN Omics Science Center from MEXT to YH; A grant of the Genome Network Project from the Ministry of Education, Culture, Sports, Science and Technology, Japan to YH (http://genomenetwork.nig.ac.jp/index_e.html). We would also like to thank all of the members in the FANTOM consortium for fruitful collaboration and cooperation in particular thanks to F. Hori for information collection and C. Wells and J. Quackenbush for discussions on the microRNA microarrays. ARRF was supported by a CJ Martin Fellowship from the Australian NHMRC (ID 428261).

Supplementary Information accompanies the paper on the Leukemia website (http://www.nature.com/leu).

# Figure legends

**Figure 1 - DNA content analysis of microRNA transfections by flow cytometry**

DNA content was assessed using propidium iodide staining 48hr post transfection. Note: mir-424 and mir-503 induced a G1 accumulation, mir-222 induced a G2 accumulation and mir-155 depleted G2 and accumulated a sub-G1 population. Untreated THP-1 cells and THP-1 transfected with scrambled RNA duplex (NegCon) are included as controls.

**Figure 2 – mir-424 and mir-503 are from a polycistronic cluster and their seed regions match the mir-16 family of microRNAs.**

**A**) Genomic organisation of hsa-mir-424-503, ESTs, mature microRNAs, TSS and RT-PCR amplicons are shown, **B**) RT-PCR confirmation of the precursor RNA (gDNA = genomic DNA positive control, RT(-) negative control RT reaction minus reverse transcriptase. Amplicon descriptions – 1: mir-424-mir-503, 2 & 3: TSS1 and TSS2 to mir-503, 4 & 5: TSS1 and TSS2 to mir-424, **C**) Multiple alignment of mir-424 and mir-503 to mir-16 family.  NOTE: qRT-PCR also confirmed a transcript spanning mir-424 and mir-503 (Supplementary Fig. 4).

**Figure 3 – Luciferase reporter assays**

Relative luciferase expression levels of 3' UTR reporter constructs transfected with synthetic precursors of A) mir-424, B) mir-503 and C) mir-9, compared to a scrambled negative control. Previously reported targets CCND1, CCNE1 and PRDM1 are marked with an asterix *. Predicted targets are shown in grey, predicted non-targets lacking sites are shown in black. Confirmed targets are shown by grey boxes.

**Figure 4 – Primary transcript expression summaries**

**A)** Expression of the mir-9-3 primary transcript by Illumina microarray (probeID 7560564). The transcript is down-regulated by PMA and also by mir-424 and mir-503 over-expression (B-statistics of 8.56 and 14.20 respectively). **B)** Expression of the putative mir-424-mir-503 primary transcript by Illumina microarray (probeID 6270392). The transcript is induced with PMA and MLL-MLLT3 knock-down (B-statistic of 0.51). Note: knock-down of a previously reported regulator SPI1(PU.1) had no effect on expression (17).

**Figure 5 - Proposed microRNA regulated quiescence in monocytic differentiation**

The PMA induced microRNAs mir-155, mir-222, mir-424 and mir-503 indirectly promote differentiation and directly down-regulate proliferation by targeting multiple cell cycle. White nodes are pro-differentiative nodes, grey nodes are anti-differentiative. Direct and indirect edges are shown as solid and dashed lines respectively.

# Tables

**Table 1 - microRNAs induced or repressed >= 3 fold in THP-1 after 96hours PMA induced differentiation.**

| Dynamically regulated microRNAs | microRNA array fold change | Small RNA sequencing | |
|---|---|---|---|
| | | fold change | max cpm |
| Induced | | | |
| miR-221 | 5.7 | 10.9 | 106251 |
| miR-24 | 3.2 | 2.2 | 65581 |
| miR-222 | 3.3 | 6.4 | 55722 |
| miR-124 | 39.3 | 44 | 30058 |
| miR-22 | 14.7 | 7.6 | 18316 |
| miR-29b | 5.3 | 3.4 | 10154 |
| miR-29a | 5.2 | 2 | 6087 |
| miR-503 | 6.8 | 8.3 | 4099 |
| miR-424 | 5.1 | 1.9 | 1971 |
| miR-155 | 3 | 1.3 | 1445 |
| miR-30a-5p | 4.3 | 1.9 | 568 |
| miR-132 | 12.3 | 11.5 | 294 |
| miR-542-3p | 4.8 | 2.6 | 172 |
| miR-137 | 17.3 | 9.5 | 153 |
| miR-151 | 10.2 | 1.8 | 105 |
| miR-133a | 7.7 | ND | 70 |
| miR-542-5p | 4.3 | ND | 67 |
| miR-146b | 5.6 | ND | 23 |
| miR-193b | 3.7 | ND | 19 |
| miR-133b | 12 | ND | < 1 |
| miR-146a | 17 | ND | ND |
| Repressed | | | |
| miR-210 | 0.3 | 0.5 | 22801 |
| miR-9 | 0.3 | 0.04 | 17871 |

microRNAs are ranked by maximum counts per million in the small RNA sequencing which is described in (25).

**Table 2 - Top 5 gene ontologies enriched in genes down-regulated with over-expression of mir-424 and mir-503**

| microRNA | p-value | term | ontology |
|---|---|---|---|
| mir-424 | 3.86E-11 | gene expression | GO:0010467 |
| | 3.39E-08 | RNA metabolic process | GO:0016070 |
| | 2.39E-07 | nucleobase, nucleoside, nucleotide and nucleic acid metabolic process | GO:0006139 |
| | 2.39E-07 | cell cycle phase | GO:0016070 |
| | 3.12E-06 | cell cycle | GO:0006139 |
| mir-503 | 1.40E-05 | cell division | GO:0051301 |
| | 1.84E-04 | cell cycle | GO:0007049 |
| | 1.84E-04 | cell cycle phase | GO:0022403 |
| | 1.84E-04 | M phase | GO:0000279 |
| | 2.34E-04 | mitosis | GO:0007067 |

Fig. 1

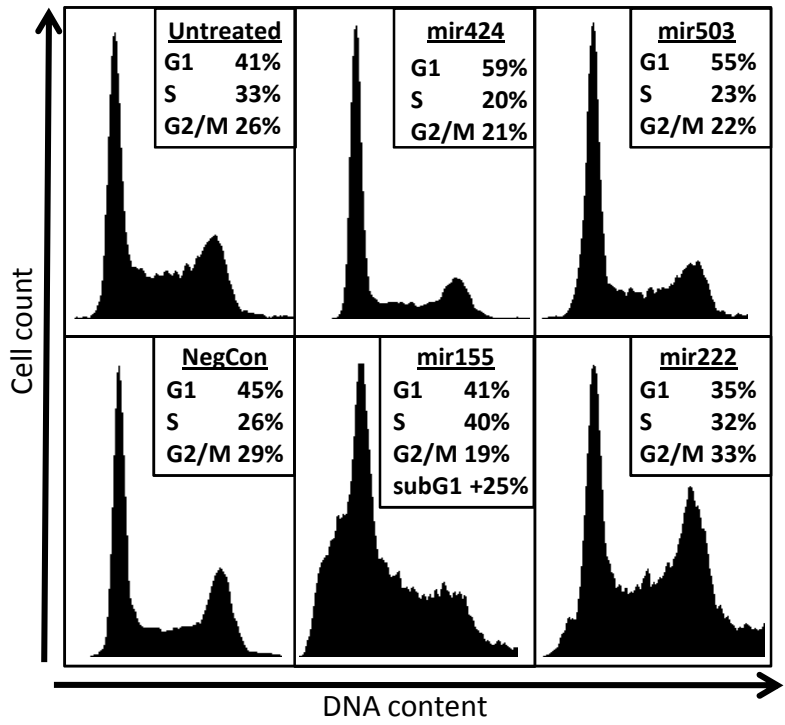

Fig. 2

A

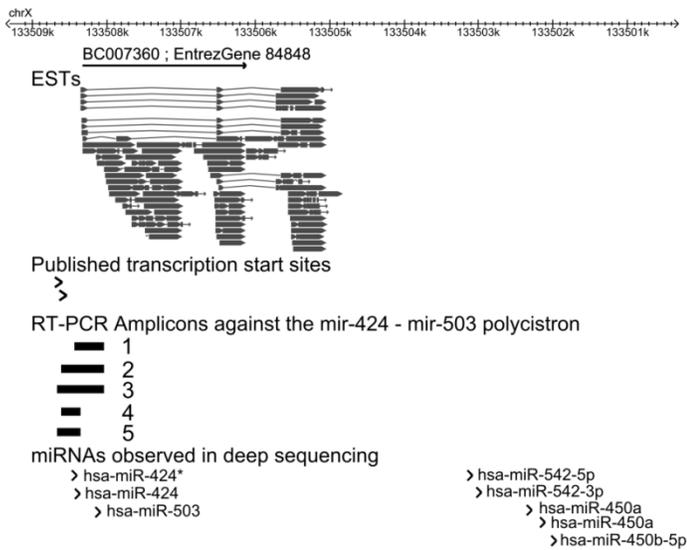

B

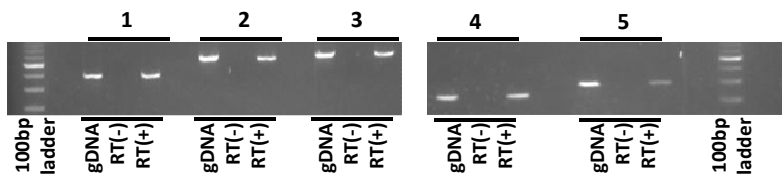

C
```
mir503      UAGCAGCGGGAACAGUUCUGCAG
mir424      CAGCAGCAAUUCAUGUUUUGAA

mir16       UAGCAGCACGUAAAUAUUGGCG
mir15A      UAGCAGCACAUAAUGGUUUGUG
mir15b      UAGCAGCACAUCAUGGUUUACA
mir195      UAGCAGCACAGAAAUAUUGGC
mir497      CAGCAGCACACUGUGGUUUGU

consensus   uAGCAGCacauaauggUuugca-
```

Fig. 3

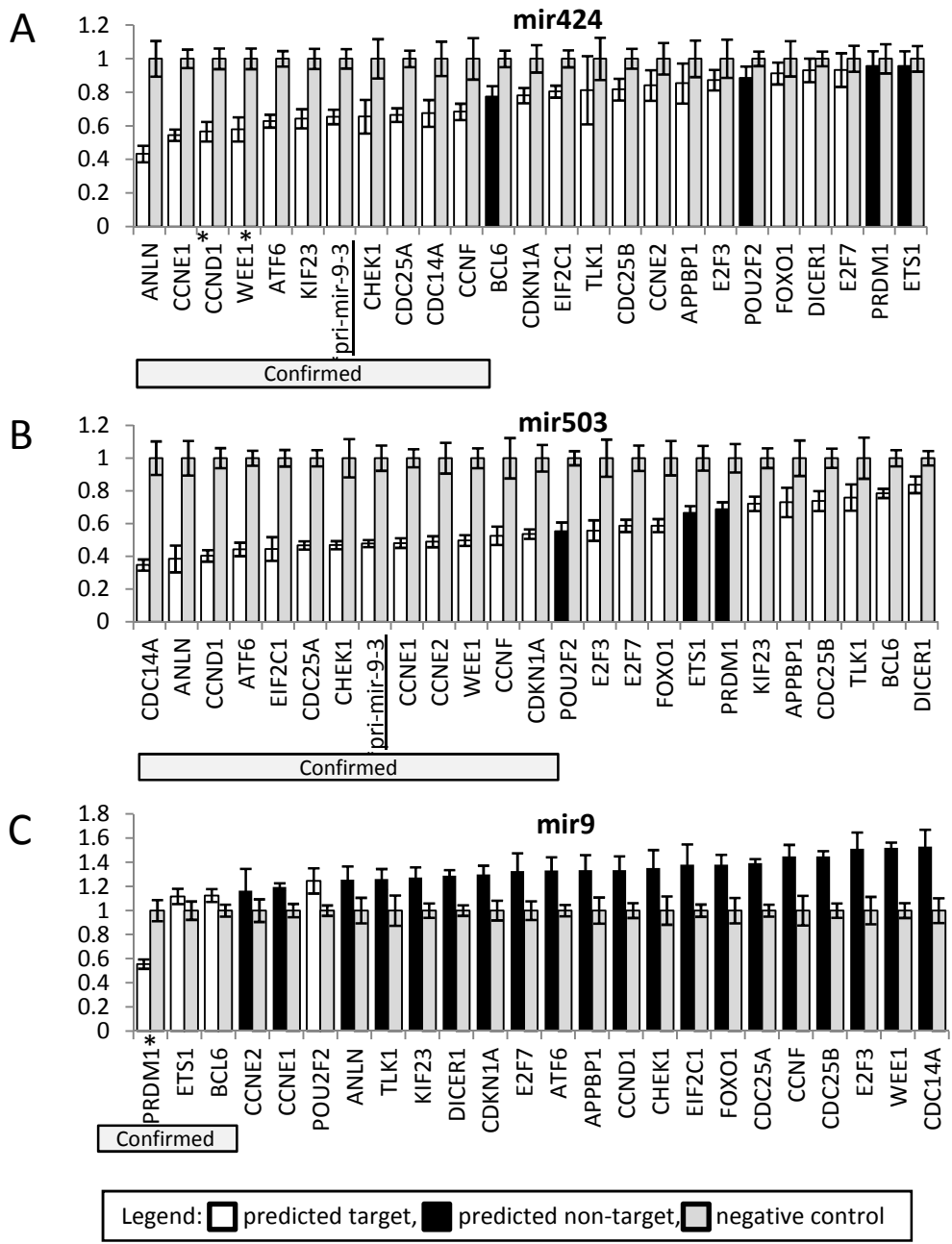

Fig. 4

A 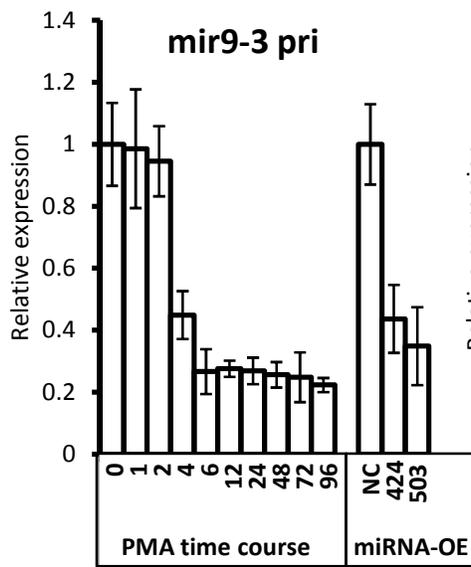

B 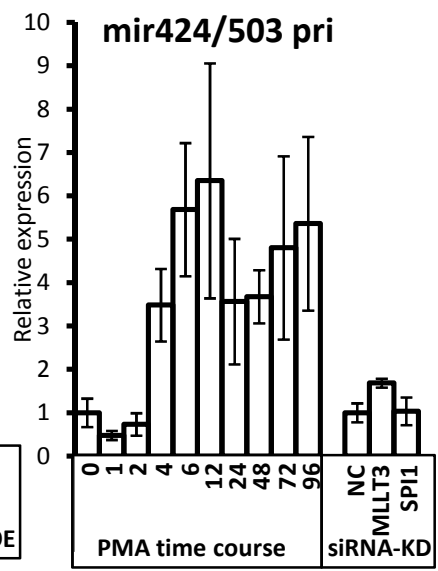

Fig. 5

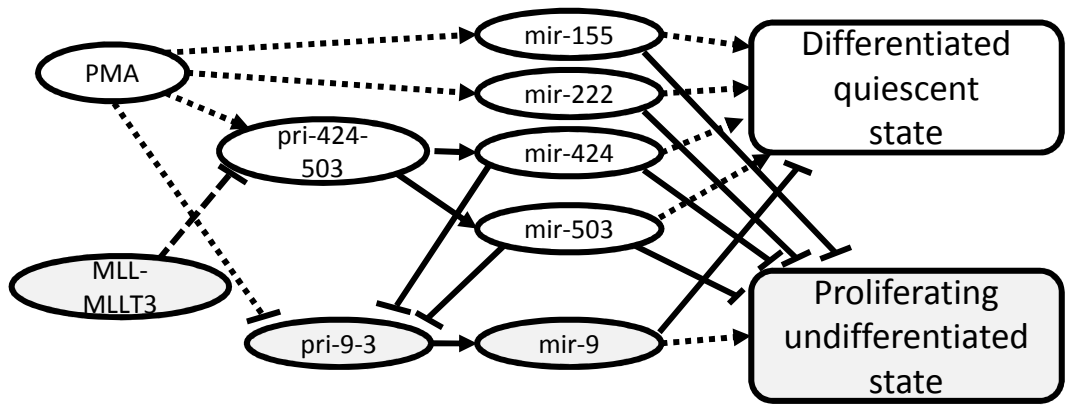

# Supplementary Materials and methods

## Cell culture and RNA extraction

Human monoblast leukemia THP-1 cells were maintained in RPMI1640 medium (Invitrogen, Carlsbad, CA, USA), 1mM sodium pyruvate, 10 mM HEPES supplemented with 0.001 % (v/v) 2-mercaptoethanol, 10 µg/ml Penicillin/ Streptomycin (Invitrogen) and 10% fetal bovine serum at 37°C in a 5% $CO_2$ and 95% $O_2$ atmosphere. RNA was purified for expression analysis by Qiagen RNeasy columns, Takara FastPure RNA Kit or TRIzol. RNA quality was checked by Nanodrop and Bioanalyser. For the pre-microRNA over-expression experiments Total RNA was extracted 48h after transfection, using the FastPure RNA kit (TAKARA BIO, Ohtsu, Shiga, Japan) in accordance with the manufacturer's instructions.

## pre-miRNA overexpression experiments

THP-1 cells were seeded in 6 cm dishes at a density of $1 \times 10^6$ cells/dish for transfection. Transfection was performed with 1.6 µg/ml (final concentration) of Lipofectamine 2000 (Invitrogen) and 20 µM (final concentration) of pre-miRNA (Ambion or Nihon-shinyaku) by reverse transfection protocol in accordance with the manufacturer's instructions. For the co-over-expression experiment 5uM for mir-155, mir-222, mir-424 and mir-503 was used.

## Agilent miRNA microarrays

RNA purification, sample preparation and hybridization to Agilent Human miRNA Microarrays (Agilent) was performed essentially as described in the Agilent Technical Manual using 100 ng total RNA. Microarrays were scanned with Agilent's DNA Microarray Scanner and expressed numerically with Agilent Feature Extraction Software. The biological triplicate data was 'per chip' normalized in GeneSpring GX (Agilent). Each measurement was divided by the $90^{th}$ percentile of all measurements in that sample. Note: due to the small number of probes on the arrays (534), per chip normalization is generally not recommended however an alternative such as spike in controls was not available, therefore, we trialed normalization to the median and other deciles. The $90^{th}$ percentile gave us the most consistent profiles between the biological replicate time-courses.

## Deep Sequencing data analysis

Library generation and sequence mapping is described previously (1). All sequences shorter than 18 nucleotides were discarded. The remaining sequences were aligned against all mature and star sequences in mirbase (release 12) using nexalign (T.Lassman manuscript in preparation). The raw expression for each miRNA sequence was determined by counting the number of sequences mapping to genome loci overlapping the genomic location of the mature miRNA. Normalized expression values were derived by dividing each raw expression with the total expression of miRNAs at each time-point and multiplying by one million.

**Illumina microarray analysis**

**Sample amplification and hybridisation**

RNA (500 ng) was amplified using the Illumina TotalPrep RNA Amplification Kit, according to manufacturer's instructions. cRNA was hybridized to Illumina Human Sentrix-6 bead chips Ver.2, according to standard Illumina protocols (http://www.illumina.com).

**Raw data curation and normalization of Illumina microarray**

Chips scans were processed using Illumina BeadScan and BeadStudio software packages and summarized data was generated in BeadStudio (version 3.1). The summarized data from BeadStudio was imported into R/BioConductor using the readBead function from the BeadExplorer package (http://bioconductor.org/packages/1.9/bioc/html/BeadExplorer.html). Background adjustment and quantile normalization (2) was performed using algorithms within the affymetrix package (3) (function: bg.adjust and normalize.quantiles). The normalized data was exported out off R/BioConductor with write.beadData function.

**Normalisation and statistical analysis of Illumina microarray data**

All microarray experiments were conducted in biological triplicate. A gene was considered detected if the average detection score (p-value) of the three replicates was less than 0.01. Quantile normalization and B-statistic calculations were carried out using the lumi and limma packages of Bioconductor in the R statistical language (4-6). For differential gene expression during the timecourse and between siRNAs, pre-miRNAs and negative control transfections we required a B-statistic ≥ 2.5, fold change ≥ 2 and the gene had to be detected in one of the conditions (average detection score ≤ 0.01).

**RT-PCR confirmation of the hsa-mir-424, hsa-mir-503 precursor**

Primers were designed overlapping the mature hsa-mir-424 and hsa-mir-503 sequences and internal to the transcription start sites described in (7). The primers for amplicons 1 to 5 are the following. amplicon 1 (chrX:133508023-133508403) used primer1: GGG ATA CAG CAG CAA TTC ATG T, and primer4: TTA CCC TGG CAG CGG AAA CAA TAC, amplicon 2 (chrX:133508023-133508586) used primer2: CAC CTG CAG CTC CTG GAA ATC AAA and primer4, amplicon 3 (chrX:133508023-133508641) used primer4 and primer3: CGT TGT TCC AAG ATT CAT CCT CAG GG, amplicon 4 (chrX:133508337-133508586) used primer2 and primer5: GGT ATA GCA GCG CCT CAC GTT T and amplicon 5 (chrX:133508336-133508641) used primer3 and primer5.

cDNA was made from a pool of THP-1 RNAs across the PMA time-course (0〜96hr 10point Mix). Templates tested by PCR were 1) positive control Genomic DNA 100ng/100ul, 2) negative control RT(-) total RNA 12.5ng/25ul, and 3) test RT(+) total RNA 12.5ng/25ul RNA).

**MicroRNA target predictions**

MicroRNA target predictions were downloaded from http://www.targetscan.org/ (version 4.2 April 2008), and http://www.mirz.unibas.ch/ElMMo2/ (release 2 January 2008).

**Luciferase reporter constructs to test sites in the 3'UTR of predicted targets**

Luciferase reporter constructs were generated from 3 fragment fusion PCR as described previously (8). The fragments consisted of 1) a CMV promoter driving a destabilised luciferase (the PCR template for this was generated by subcloning the destabilised luciferase from pGL4.12_luc2CP (Promega) into pMIR-REPORT (Ambion)), 2) A fragment containing the predicted miRNA target sites in 3'UTRs of candidate target mRNAs to be tested (PCR primers are given in **supplementary table 3**). 3) an SV40 late poly-adenylation signal amplified from pG5 vector (Promega).

Samples for the assay were constructed by using PCR. The followings are PCR primers used; target-specific forward primer, 5'-GAA GGA GCC GCC ACC ATG-3' followed by approximately 20-base 5'-end target sequence; target-specific reverse primer, 5'-CAA TTT CAC ACA GGA AAC TCA-3' followed by approximately 20-base 3'-end target sequence; PMIR_CMV_F, 5'-GGG TCA TTA GTT CAT AGC CCA-3'; Luc2CP_Luc_R, 5'-CAT GGT GGC GGC TCC TTC TTA GAC GTT GAT CCT GGC

GCT-3'; FSV40LPAS02, 5'-GTT CCT GTG TGA AAA TTG TTA TCC GCT GCA GAC ATG ATA AGA TAC ATT G-3'; RSV40LPAS01, 5'-AGC AAG TTC AGC CTG GTT AAG ATC CTT ATC GAT TTT ACC AC-3'; PMIR_CMV_Nested_F, 5'-GGG AGG AGA AGC ATG AAT TCA AGG-3'; LGT10L, 5'-AGC AAG TTC AGC CTG GTT AAG-3'; FPCMV5, 5'-GCC ATG TTG GCA TTG ATT ATT GAC-3'; (The sequences in bold type are complementary to the common tag sequences of the target-specific forward and reverse primers, respectively).

Each target sequence was amplified with the designed target-specific forward and reverse primers. The fragment for the CMV promoter and the Luc2CP gene was PCR-amplified from pMIR_Luc2CP vector using the primer sets PMIR_CMV_F and Luc2CP_Luc_R. The fragment for the SV40 late poly-adenylation signal (SV40LPAS) was PCR-amplified from pG5 vector (Promega) using the primer sets FSV40LPAS02 and RSV40LPAS01. Overlapping PCR was carried out to connect the target fragments with the CMV_Luc2CP-fragment and the SV40LPAS fragment using the primer pair PMIR_CMV_Nested_F and LGT10L. PCR conditions were based on those in our previous reports.

The fragment for the CMV promoter, Renilla luciferase gene and SV40LPAS was also PCR-amplified from pRL-CMV using the primer sets FPCMV5 and RSV40LPAS01. All PCR products were checked by agarose gel electrophoresis after the amplification.

CR Primers were designed flanking miRNA target sites in 3'UTRs of candidate target mRNAs (supplementary table 1). A CMV-luciferase PCR fragment Fusion PCR was used to generate constructs.

**MicroRNA target validation by luciferase assay**

miRNA target reporter assays were carried out in 96-well assay plates and were repeated three times. Each target construct (0.2 µl), the Renilla luciferase fragment (0.2 µl) and pre-miRNA (2 pmol) were diluted to 95 µl with Opti-MEM (Invitrogen) followed by mixing 50 µl of 2% Lipofectamine2000 (Invitrogen). After 20 min incubation at room temperature, the suspended Hela cells (88,000 cells/55 µl) were mixed for transfection (total 200 µl). After 48h of incubation, fire fly and Renilla luciferase reporters were sequentially measured using the Dual-Glo Luciferase Assay system (Promega) according to the manufacture's protocol. The reporter activity for fire fly luciferase was normalized by that for Renilla luciferase.

## Flow cytometry

48 hours after pre-miRNA transfection, cells were harvested and fixed in 70% ince cold ethanol. Fixed cells were washed in PBS and then DNA stained (2.5 µg/ml propidium iodide, and 0.5 mg/ml RNase A in PBS). Flow cytometry analysis was done using a FACS Calibur (Becton Dickinson, Franklin, NJ, USA) instrument following the manufacturer's recommended protocol. Data were collected and processed using the FlowJo FACS analysis software (Tree Star, Inc., Ashland, OR, USA).

## Data deposition

All expression data has been deposited at the <u>C</u>enter for <u>I</u>nformation <u>B</u>iology gene <u>EX</u>pression database (CIBEX (9)). Accession numbers for the Illumina mRNA measurements are CBX46 (PMA time-course), CBX45 (pre-miRNA over-expression), and CBX47 (52 transcription factor knockdowns). The accession number for the Agilent miRNA array experiment is CBX49. The small RNA sequences of Taft et al. (1) are deposited at DDBJ (Accessions: AIAAF0000001 - AIAAF0055261, AIAAG0000001 - AIAAG0013956, AIAAH0000001 - AIAAH0046376, AIAAI0000001 - AIAAI0087752, AIAAJ0000001 - AIAAJ0100819, AIAAP0000001 - AIAAP0061402, AIAAQ0000001 - AIAAQ0033032, AIAAR0000001 - AIAAR0049623, AIAAS0000001 - AIAAS0055455, AIAAT0000001 - AIAAT0061351).

## Authors' contributions statement

MK carried out the Illumina mRNA and Agilent miRNA array experiments. Y. Tomaru performed the pre-miRNA over-expression experiments. NN performed the luciferase reporter experiments and RT-PCR confirmation of the mir-424-mir-503 pri-miRNA. Y. Takahashi performed the FACs experiments. TL conceptualized some of the bioinformatic analysis. MdH provided the normalised miRNA tag counts from the small RNA libraries. AK was involved in experimental testing. MS, JY, were involved in supervision and general discussions. TL, JK, HS, YH and DAH were involved in conceptualization of FANTOM4 and the choice of the THP-1 model. ARRF designed and analysed most of the experiments (including the array profiling, over-expression, reporter assays and FACs experiments), carried out the bioinformatic analysis and wrote the manuscript.

**References for supplementary information**

# Supplementary legends

# Supplementary figures:

## Supplementary Figure 1 - Monocytic markers induced with microRNA over-expression

The mRNA expression profiles of monocytic markers induced with PMA and upon over-expression of the four pro-differentiative microRNAs (hsa-mir-424, hsa-mir-503, hsa-mir-222, hsa-mir-155). Illumina array expression is shown relative to 0h for the time-course and relative to the scrambled negative control transfection for the microRNA over-expression experiments. The published PMA time course data was extracted from FANTOM4 EdgeExpressDB (22).

## Supplementary Figure 2 – Combinatorial effect of four miRNA co-transfection

A) Modest induction of CD14 by the four miRNA mix.

B) Pro-differentiative changes induced by individual miRNAs and from the four way mix.

## Supplementary Figure 3 - The mRNAs of predicted microRNA targets are down-regulated

Histograms showing the median fold change in mRNA levels of cells transfected with synthetic pre-microRNAs relative to negative control transfections for predicted microRNA targets (red) or all genes (black), and for varying strengths of predictions (blue bars). Both TargetScan (1) and EIMMO (2) predictions are shown. Bin 0 corresponds to genes with prediction strengths between >0 and <0.1, Bin 0.1 to values between >0.1 and <0.2 etc. Bin 1 is only used for

EIMMO (2) predictions with values greater than 1. TS_all and TS_cons correspond to all TargetScan (1) predictions versus conserved only predictions. Prediction strength for TargetScan corresponds to context-score as defined by the TS paper. Note mir-142 (the microRNA that does not change levels during the differentiation has the smallest change).

## Supplementary Figure 4 – qRT confirmation of mir-424-mir-503 polycistron

qRT-PCR confirmation of the transcript spanning mir424 and mir503. NOTE: RT- samples are not detected (ct=40), indicating no DNA contamination.

## Supplementary Figure 5 – mir-424-mir-503 target site in pri-mir-9-3

Location of a potential mir-424-mir-503 target site in the primary transcript of mir-9-3. Note: A luciferase reporter construct containing this UTR was significantly down-regulated by both mir-424 and mir-503 over-expression.

# Supplementary tables:

## Supplementary table 1 – Perturbed genes

Combined list of genes significantly perturbed in any of the microRNA over-expression experiments.

## Supplementary Table 2 - Pro-differentiative effects of pre-microRNA transfections

microRNAs that induce changes in mRNA expression similar to those seen in PMA treated THP-1 cells are shown in gold, and microRNAs that repress changes seen with PMA are shown in green. The label diff corresponds to the fraction of changes that match the PMA differentiative response (ie sum of changes that upregulate genes that are also upregulated with PMA plus the changes that downregulated genes that are also down-regulated with PMA). The label undiff is the opposite (ie. The microRNA induced expression changes in the opposite direction to PMA induced expression changes).

## Supplementary table 3 – Gene ontology enrichment

Gene ontology enrichments for genes up or down-regulated in THP-1 miRNA over-expression experiments.

## Supplementary table 4 – predicted direct targets down-regulated at the RNA level

A list of predicted targets of all microRNAs tested that are supported by mRNA down-regulation (at least 0.75x that of control).

## Supplementary table 5 – PCR primers used for 3'UTR luciferase constructs

Fold change of targets tested by luciferase assay. PCR primers and amplicons sizes are included.

**Supplementary Table 2 - Pro-differentiative effects of pre-microRNA transfections**

microRNAs that induce changes in mRNA expression similar to those seen in PMA treated THP-1 cells are shown in gold, and microRNAs that repress changes seen with PMA are shown in green. The label diff corresponds to the fraction of changes that match the PMA differentiative response (ie sum of changes that upregulate genes that are also upregulated with PMA plus the changes that downregulated genes that are also down-regulated with PMA). The label undiff is the opposite (ie. The microRNA induced expression changes in the opposite direction to PMA induced expression changes).

| Micro-RNA | Down-regulated with microRNA | | | | | Up-regulated with microRNA | | | | | ALL changes | | | diff/undiff |
|---|---|---|---|---|---|---|---|---|---|---|---|---|---|---|
| | Number of probes | Down in PMA | Up in PMA | Fraction down in PMA | Fraction up in PMA | Number of probes | Up in PMA | Down in PMA | Fraction up in PMA | Fraction down in PMA | Number of probes | Fraction matches PMA (diff) | Fraction opposite to PMA (undiff) | |
| mir-155 | 178 | 19 | 12 | 10.7% | 6.7% | 340 | 65 | 17 | 19.1% | 5.0% | 518 | 16.2% | 5.6% | 2.9 |
| mir-503 | 735 | 75 | 15 | 10.2% | 2.0% | 244 | 35 | 23 | 14.3% | 9.4% | 979 | 11.2% | 3.9% | 2.9 |
| mir-424 | 88 | 4 | 11 | 4.5% | 12.5% | 121 | 31 | 7 | 25.6% | 5.8% | 209 | 16.7% | 8.6% | 1.9 |
| mir-222 | 109 | 15 | 12 | 13.8% | 11.0% | 96 | 12 | 5 | 12.5% | 5.2% | 205 | 13.2% | 8.3% | 1.6 |
| mir-132 | 148 | 15 | 23 | 10.1% | 15.5% | 150 | 12 | 5 | 8.0% | 3.3% | 298 | 9.1% | 9.4% | 1.0 |
| mir-221 | 113 | 9 | 19 | 8.0% | 16.8% | 69 | 5 | 6 | 7.2% | 8.7% | 182 | 7.7% | 13.7% | 0.6 |
| mir-142 | 376 | 21 | 92 | 5.6% | 24.5% | 217 | 10 | 21 | 4.6% | 9.7% | 593 | 5.2% | 19.1% | 0.3 |
| mir-22 | 132 | 6 | 41 | 4.5% | 31.1% | 65 | 3 | 6 | 4.6% | 9.2% | 197 | 4.6% | 23.9% | 0.2 |
| mir-29a | 232 | 9 | 57 | 3.9% | 24.6% | 116 | 6 | 23 | 5.2% | 19.8% | 348 | 4.3% | 23.0% | 0.2 |
| mir-9 | 145 | 5 | 52 | 3.4% | 35.9% | 119 | 6 | 16 | 5.0% | 13.4% | 264 | 4.2% | 25.8% | 0.2 |
| mir-29b | 227 | 7 | 63 | 3.1% | 27.8% | 107 | 6 | 19 | 5.6% | 17.8% | 334 | 3.9% | 24.6% | 0.2 |
| mir-146b | 173 | 3 | 51 | 1.7% | 29.5% | 64 | 3 | 14 | 4.7% | 21.9% | 237 | 2.5% | 27.4% | 0.1 |
| **mix** | 69 | 12 | 6 | 17.4% | 8.7% | 209 | 57 | 5 | 27.3% | 2.4% | 278 | 24.8% | 4.0% | 6.3 |
| | | median | | 5.1% | 20.6% | | | | 6.4% | 9.3% | | 6.5% | 16.4% | |
| | | std_dev | | 3.8% | 10.6% | | | | 6.9% | 6.1% | | 5.0% | 8.7% | |
| | | Threshold | | 8.9% | 31.2% | | | | 13.3% | 15.5% | | 11.5% | 25.1% | |

# Supplementary Figure 1 - Monocytic markers induced with microRNA over-expression

The mRNA expression profiles of monocytic markers induced with PMA and upon over-expression of the four pro-differentiative microRNAs (hsa-mir-424, hsa-mir-503, hsa-mir-222, hsa-mir-155). Illumina array expression is shown relative to 0h for the time-course and relative to the scrambled negative control transfection for the microRNA over-expression experiments. The published PMA time course data was extracted from FANTOM4 EdgeExpressDB (22).

**Supplementary Figure 2 – Combinatorial effect of four miRNA co-transfection**
A) Modest induction of CD14 by the four miRNA mix.
B) Pro-differentiative changes induced by individual miRNAs and from the four way mix.

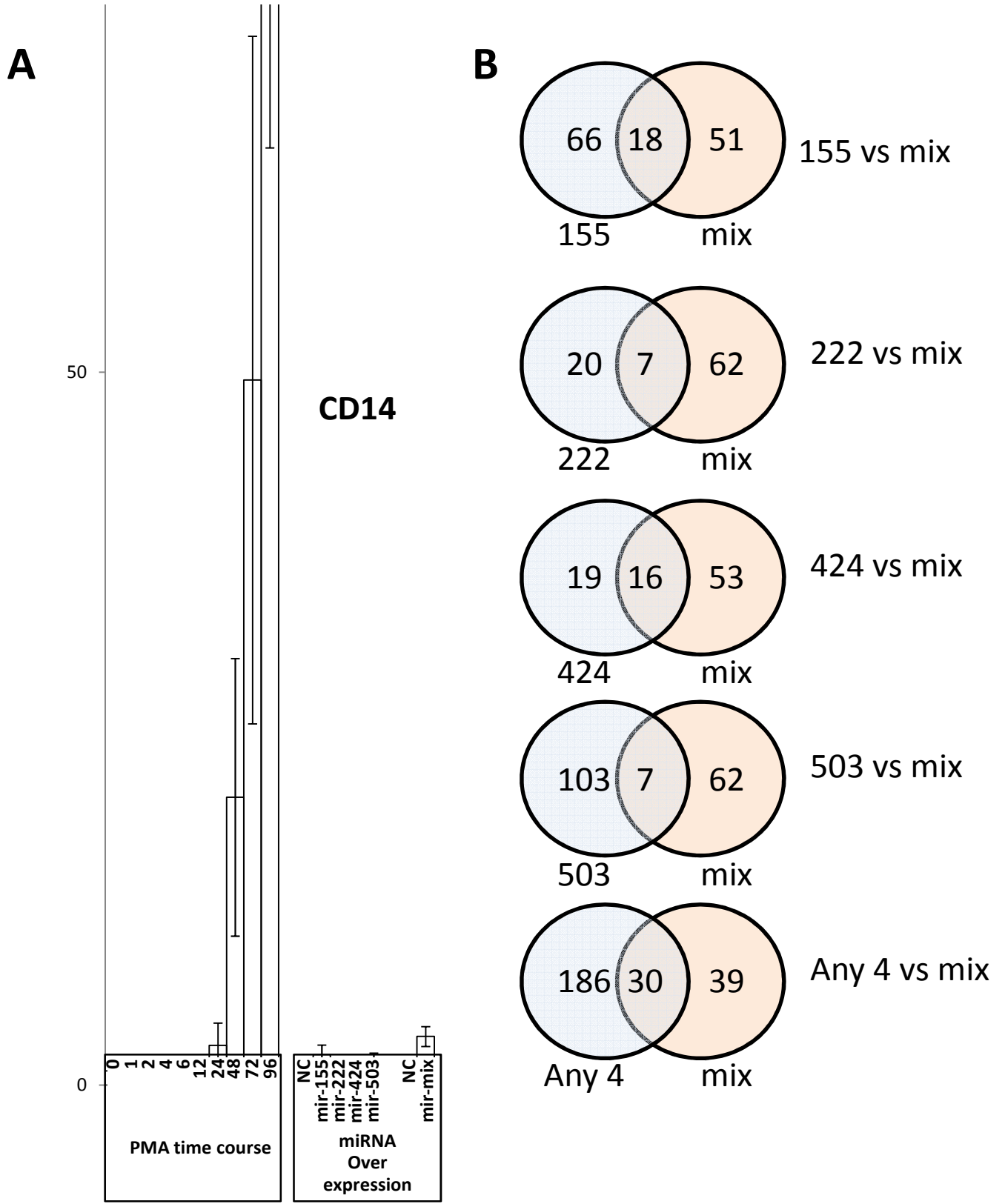

# Supplementary Figure 3 - The mRNAs of predicted microRNA targets are down-regulated

Histograms showing the median fold change in mRNA levels of cells transfected with synthetic pre-microRNAs relative to negative control transfections for predicted microRNA targets (red) or all genes (black), and for varying strengths of predictions (blue bars). Both TargetScan (1) and EIMMO (2) predictions are shown. Bin 0 corresponds to genes with prediction strengths between >0 and <0.1, Bin 0.1 to values between >0.1 and <0.2 etc. Bin 1 is only used for EIMMO (2) predictions with values greater than 1. TS_all and TS_cons correspond to all TargetScan (1) predictions versus conserved only predictions. Prediction strength for TargetScan corresponds to context-score as defined by the TS paper. Note mir-142 (the microRNA that does not change levels during the differentiation has the smallest change).

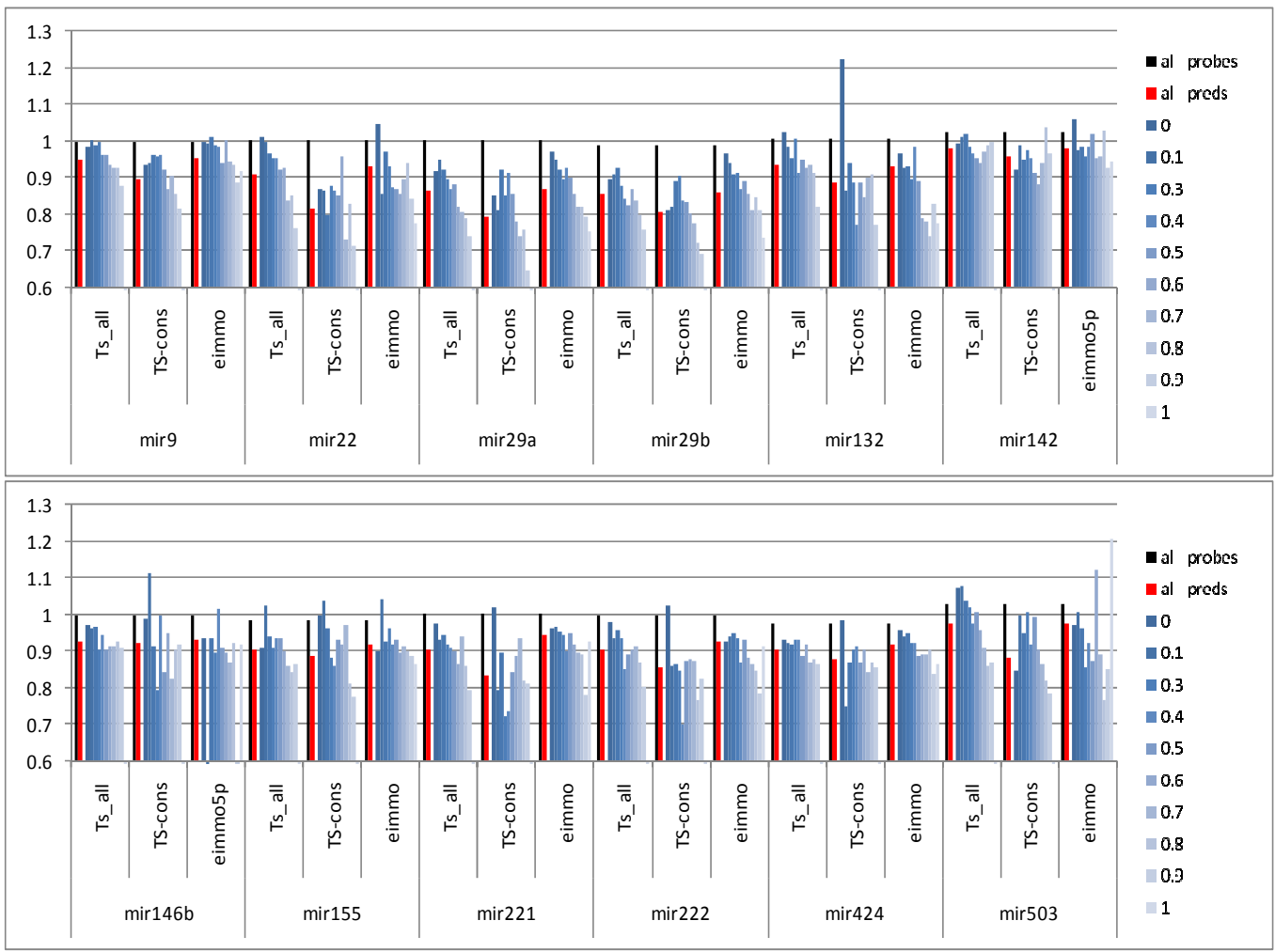

# Supplementary Figure 4 – qRT confirmation of mir-424-mir-503 polycistron

qRT-PCR confirmation of the transcript spanning mir424 and mir503. NOTE: RT(-) samples are not detected (ct=40), indicating no DNA contamination.

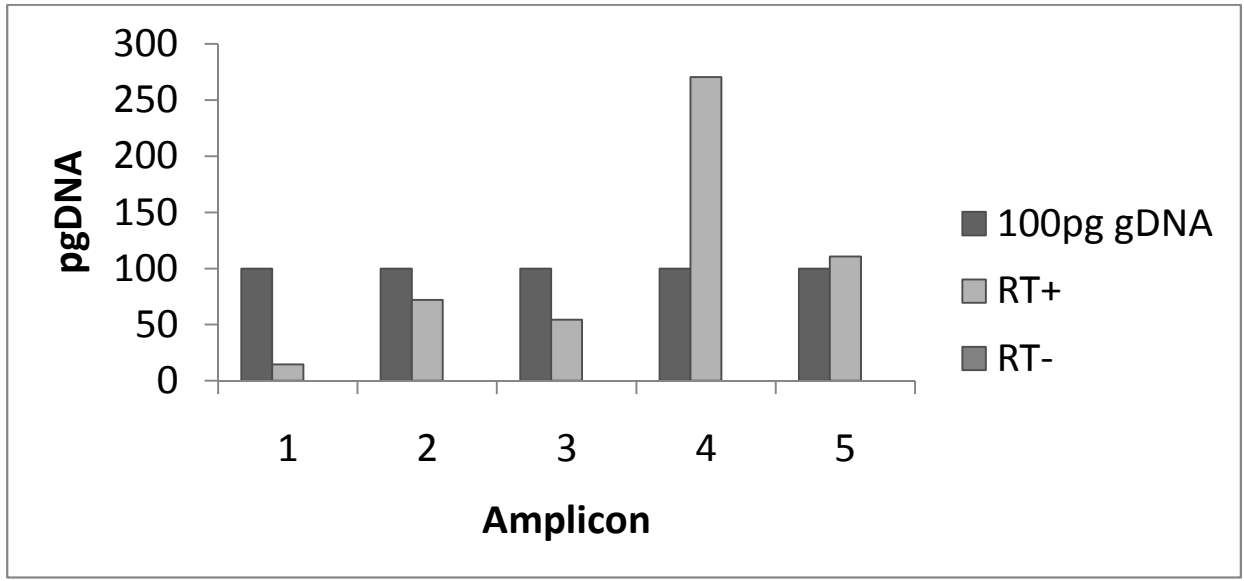

|  | Primer_1+4 | | Primer_2+4 | | Primer_3+4 | | Primer_2+5 | | Primer_3+5 | |
|---|---|---|---|---|---|---|---|---|---|---|
| Human Genomic DNA 0.1/well | rep1 32.7 | | rep1 39.6 | | rep1 36.6 | | rep1 32.3 | | rep1 33.2 | |
| | rep2 34.3 | 34.2 | rep2 37.9 | 38.2 | rep2 36.9 | 37.6 | rep2 33.4 | 33.0 | rep2 31.7 | 32.9 |
| | rep3 35.5 | | rep3 37.0 | | rep3 39.4 | | rep3 33.2 | | rep3 33.7 | |
| THP1_timecourse_pool_RT(+) total RNA 2.5ng/well | rep1 37.4 | | rep1 37.5 | | rep1 38.3 | | rep1 31.7 | | rep1 31.9 | |
| | rep2 35.2 | 36.9 | rep2 40.0 | 38.6 | rep2 39.1 | 38.5 | rep2 31.4 | 31.5 | rep2 31.3 | 31.8 |
| | rep3 38.2 | | rep3 38.4 | | rep3 38.2 | | rep3 31.5 | | rep3 32.2 | |
| THP1_timecourse_pool _RT(-) total RNA 2.5ng/well | rep1 40.0 | | rep1 40.0 | | rep1 40.0 | | rep1 40.0 | | rep1 40.0 | |
| | rep2 40.0 | 40.0 | rep2 40.0 | 40.0 | rep2 40.0 | 40.0 | rep2 40.0 | 40.0 | rep2 40.0 | 40.0 |
| | rep3 40.0 | | rep3 40.0 | | rep3 40.0 | | rep3 40.0 | | rep3 40.0 | |

| Amplicons: | FWD | REV | size | Genomic span |
|---|---|---|---|---|
| 1: mir424_TO_mir503 | primer1 | primer4 | 381 | chrX:133508023-133508403 |
| 2: TSS+100_TO_mir503 | primer2 | primer4 | 564 | chrX:133508023-133508586 |
| 3: TSS2_TO_mir503 | primer3 | primer4 | 619 | chrX:133508023-133508641 |
| 4: TSS+100_TO_mir424 | primer2 | primer5 | 250 | chrX:133508337-133508586 |
| 5: TSS2_TO_mir424 | primer3 | primer5 | 305 | chrX:133508336-133508641 |

| primer1 | GGGATACAGCAGCAATTCATGT |
| primer2 | CACCTGCAGCTCCTGGAAATCAAA |
| primer3 | CGTTGTTCCAAGATTCATCCTCAGGG |
| primer4 | TTACCCTGGCAGCGGAAACAATAC |
| primer5 | GGTATAGCAGCGCCTCACGTTT |

# Supplementary Figure 5 – mir-424-mir-503 target site in pri-mir-9-3

Location of a potential mir-424-mir-503 target site in the primary transcript of mir-9-3. Note: A luciferase reporter construct containing this UTR was significantly down-regulated by both mir-424 and mir-503 over-expression.

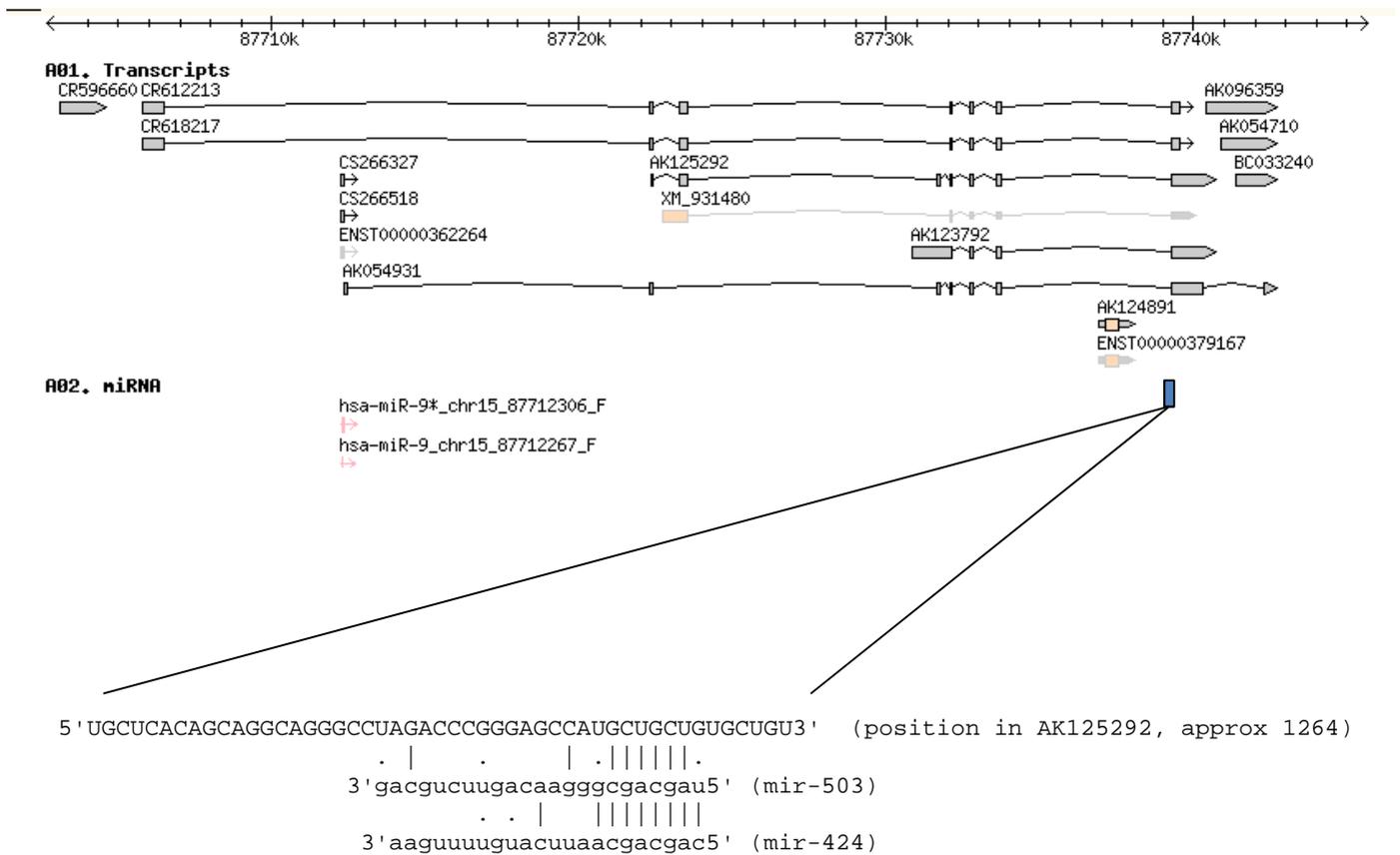

```
5'UGCUCACAGCAGGCAGGGCCUAGACCCGGGAGCCAUGCUGCUGUGCUGU3'   (position in AK125292, approx 1264)
                  . |    .      | .||||||.
                   3'gacgucuugacaagggcgacgau5' (mir-503)
                         . . |   |||||||||
                    3'aaguuuuguacuuaacgacgac5' (mir-424)

Note: Watson-crick base pairing marked with |
      G-U base pairing is marked with .
```